\documentclass[nofootinbib,pre,twocolumn,showpacs,showkeys,groupaddress,preprintnumbers,floatfix]{revtex4-1}

\usepackage{dynlearn}
\usepackage{physics}
\usepackage{multibib}
\newcites{SM}{SM References}
\usepackage{mathrsfs}
\usepackage[english]{babel}
\usepackage{amssymb,amscd}
\newtheorem{theorem}{Theorem}

\newcommand{\MV}[1]{\left\langle #1 \right\rangle}

\newcommand{\intP}{\int_{\mathcal{P}(\mathcal{H})} \!\!\!\!\!\!\!\!\!}

\def\tbf #1 {\textbf{#1} }

\begin{document}

\def\ourTitle{Beyond Density Matrices: Geometric Quantum States
}

\def\ourAbstract{A quantum system's state is identified with a density matrix. Though their
probabilistic interpretation is rooted in ensemble theory, density matrices
embody a known shortcoming. They do not completely express an ensemble's
physical realization. Conveniently, when working only with the statistical
outcomes of projective and positive operator-valued measurements this is not a
hindrance. To track ensemble realizations and so remove the shortcoming, we
explore geometric quantum states and explain their physical significance. We
emphasize two main consequences: one in quantum state manipulation and one
in quantum thermodynamics.
}

\def\ourKeywords{Quantum Mechanics, Geometric Quantum Mechanics
}

\hypersetup{
  pdfauthor={Fabio Anza},
  pdftitle={\ourTitle},
  pdfsubject={\ourAbstract},
  pdfkeywords={\ourKeywords},
  pdfproducer={},
  pdfcreator={}
}

\title{\ourTitle}

\author{Fabio Anza}
\email{fanza@ucdavis.edu}

\author{James P. Crutchfield}
\email{chaos@ucdavis.edu}

\affiliation{Complexity Sciences Center and Physics Department,
University of California at Davis, One Shields Avenue, Davis, CA 95616}

\date{\today}
\bibliographystyle{unsrt}

\begin{abstract}
\ourAbstract
\end{abstract}

\keywords{\ourKeywords}

\pacs{
05.45.-a  89.75.Kd  89.70.+c  05.45.Tp  }

\preprint{\arxiv{2008.XXXXX}}

\date{\today}
\maketitle

\paragraph*{Introduction.} Dynamical systems theory describes long-term
recurrent behavior via a system's \emph{attractors}: stable
dynamically-invariant sets. Said simply there are regions of state
space---points, curves, smooth manifolds, or fractals---the system repeatedly
visits. These objects are implicitly determined by the underlying equations of
motion and are modeled as probability distributions (measures) on the system's
state space.

Building on this, the following introduces tools aimed at studying attractors
for quantum systems. This requires developing a more fundamental concept of
``state of a quantum system'', essentially moving beyond the standard notion of
density matrices, though they can be directly recovered. We call these objects
the system's \emph{geometric quantum states} and, paralleling the
Sinai-Bowen-Ruelle measures of dynamical systems theory \cite{Eckmann1985},
they are specified by a probability distribution on the manifold of quantum
states.

Quantum mechanics is firmly grounded in a vector formalism in which states
$\ket{\psi}$ are elements of a complex Hilbert space $\mathcal{H}$. These are
the system's \emph{pure states}, as opposed to \emph{mixed states} that account
for incomplete knowledge of a system's actual state. To account for both, one
employs \emph{density matrices} $\rho$. These are operators in $\mathcal{H}$
that are positive semi-definite $\rho \geq 0$, self-adjoint $\rho
=\rho^\dagger$, and normalized $\Tr \rho = 1$.

The interpretation of a density matrix as a system's \emph{probabilistic state}
is given by \emph{ensemble theory} \cite{Pathria2011,Greiner1995}. Accordingly,
since a density matrix always decomposes into eigenvalues $\lambda_i$ and
eigenvectors $\ket{\lambda_i}$:
\begin{align}
\rho = \sum_i \lambda_i \ket{\lambda_i}\bra{\lambda_i}
  ~,
\label{eq:DiadDecomp}
\end{align}
one interprets $\rho$ as an ensemble of pure states---the eigenvectors---in
which $\lambda_i$ is the probability of an observer interacting with
state $\ket{\lambda_i}$.

However, this interpretation is problematic: It is not unique. One can write
the same $\rho$ using different decompositions, for example in terms of
$\{ \ket{\psi_k}\} \neq \{\ket{\lambda_i}\}$:
\begin{align*}
\rho = \sum_k p_k \ket{\psi_k}\bra{\psi_k}
  ~.
\end{align*}
Given the interpretation, all the decompositions identify the same quantum
state $\rho$. While one often prefers Eq. (\ref{eq:DiadDecomp})'s diagonal
decomposition in terms of eigenvalues and eigenvectors, it is not the only one
possible. More tellingly, in principle, there is no experimental reason to
prefer it to others. In quantum mechanics, this fact is often addressed by
declaring density matrices with the same \emph{barycenter} equal. A familiar
example of this degeneracy is that the maximally mixed state ($\rho \propto
\mathbb{I}$) has an infinite number of identical decompositions, each possibly
representing a physically-distinct ensemble.

Moreover, it is rather straightforward to construct systems that, despite
having the same density matrix, are in different states. For example, consider
two distinct state-preparation protocols. In one case, prepare states
$\{\ket{0},\ket{1}\}$ each with probability $1/2$; while, in the other, always
prepare states $\{\ket{-},\ket{+}\}$ each with probability $1/2$. They are
described by the same $\rho$. A complete and unambiguous mathematical concept
of state should not conflate distinct physical configurations. Not only
do such ambiguities lead to misapprehending fundamental mechanisms, they also
lead one to ascribe complexity where there is none.

Here we argue that an alternative---the \emph{geometric formalism}---together
with an appropriately adapted measure theory cleanly separates the primary
concept of a \emph{system state} from the derived concept of a density matrix
as the set of all \emph{positive operator-valued measurement statistics} generated by a system.

With this perspective in mind, we introduce a more incisive description of pure-state ensembles. 
The following argues that \emph{geometric quantum mechanics} (GQM), through its notion that 
geometric quantum states are \emph{continuous mixed states}, resolves the ambiguities. First, 
we introduce GQM. Second, we discuss how it relates to the density matrix formalism. Then we 
analyze two broad settings in which the geometric formalism arises quite naturally: quantum state 
manipulation \cite{Anza20c} and quantum thermodynamics \cite{Anza20b}. After discussing the results, 
we draw out several consequences.

\paragraph*{Geometric quantum mechanics.}
\label{sec:GQM}
References
\cite{STROCCHI1966,Kibble1979,Heslot1985,Gibbons1992,Ashtekar1995,Ashtekar1999,Brody2001,Bengtsson2017,Carinena2007,Chruscinski2006,Marmo2010,Avron2020,Pastorello2015,Pastorello2015a,Pastorello2016,Clemente-Gallardo2013}
give a comprehensive introduction to GQM. Here, we briefly summarize only the
elements we need, working with Hilbert spaces $\mathcal{H}$ of finite dimension $D$.

Pure states are points in the complex projective manifold $\mathcal{P}\left(
\mathcal{H} \right)=\mathbb{C}\mathrm{P}^{D-1}$. Therefore, given an arbitrary
basis $\left\{\ket{e_\alpha} \right\}_{\alpha=0}^{D-1}$, a pure state $\psi$ is
parametrized by $D$ complex homogeneous coordinates $Z = \left\{      Z^\alpha\right\}$, up to
normalization and an overall phase:
\begin{align*}
\ket{\psi} = \sum_{\alpha=0}^{D-1} Z^\alpha \ket{e_\alpha}
  ~,
\end{align*}
where $Z \in \mathbb{C}^{D}$, $Z \sim \lambda Z$, and $\lambda \in
\mathbb{C}/\left\{ 0\right\}$. If the system consists of a single qubit, for
example, one can always use amplitude-phase coordinates $Z = (\sqrt{1-p},\sqrt{p} e^{i\nu})$.

An \emph{observable} is a quadratic real function $\mathcal{O}(Z) \in
\mathbb{R}$ that associates to each point  $Z \in \mathcal{P}(\mathcal{H})$ the
expectation value $\bra{\psi} \mathcal{O} \ket{\psi}$ of the corresponding
operator $\mathcal{O}$ on state $\ket{\psi}$ with coordinates $Z$:
\begin{align}
\mathcal{O}(Z) = \sum_{\alpha,\beta} \mathcal{O}_{\alpha,\beta}Z^\alpha \overline{Z}^\beta
  ~,
\label{eq:GQM_Observable}
\end{align}
where $\mathcal{O}_{\alpha \beta}$ is Hermitian $\mathcal{O}_{\beta,\alpha} = \overline{\mathcal{O}}_{\alpha,\beta}$.

Measurement outcome probabilities are determined by \emph{positive
operator-valued measurements} (POVMs) $\left\{E_j\right\}_{j=1}^n$ applied to a
state \cite{Nielsen2010,Heinosaari2012}. They are nonnegative operators
$E_j\geq 0$, called \emph{effects}, that sum up to the identity: $\sum_{j=1}^n
E_j = \mathbb{I}$. In GQM they consist of nonnegative real functions $E_j(Z)\ge
0$ on $\mathcal{P}(\mathcal{H})$ whose sum is always unity:
\begin{align}
E_j(Z) = \sum_{\alpha,\beta}
  \left(E_j\right)_{\alpha,\beta} Z^\alpha \overline{Z}^\beta
  ~,
\label{eq:GQM_POVMs}
\end{align}
where $\sum_{j=1}^{n}E_j(Z) = 1$.

Complex projective spaces, such as $\mathcal{P}(\mathcal{H})$, have a preferred
metric $g_{FS}$---the \emph{Fubini-Study metric} \cite{Bengtsson2017}---and an
associated volume element $dV_{FS}$ that is coordinate-independent and
invariant under unitary transformations. The geometric derivation of $dV_{FS}$
is beyond our immediate goals here. That said, it is sufficient to give its
explicit form in the ``probability + phase'' coordinate system $Z^{\alpha} =
\sqrt{p_\alpha}e^{i\nu_\alpha}$ that we use for explicit calculations: 
\begin{align*}
dV_{FS}
  & = \sqrt{\det g_{FS}}
  \prod_{\alpha=0}^{D-1} dZ^\alpha d\overline{Z}^\alpha \\
  & =  \prod_{\alpha=1}^{D-1} \frac{dp_\alpha d\nu_\alpha}{2}
  ~.
\end{align*}
Notice how $p_0$ and $\nu_0$ are not involved. This is due to
$\mathcal{P}(\mathcal{H})$'s projective nature which guarantees that we can
choose a coordinate patch in which $p_0 = 1 - \sum_{\alpha=1}^{D-1}p_\alpha$
and $\nu_0 = 0$.

\paragraph*{Geometric quantum states.}
This framework makes it very natural to view a quantum state as a functional
encoding that associates expectation values to observables, paralleling the
$C^{*}$-algebras formulation of quantum mechanics \cite{Strocchi2008a}. Thus,
states are described as functionals $P[\mathcal{O}]$ from the algebra of
observables $\mathcal{A}$ to the real line: 
\begin{align}
P_q[\mathcal{O}]
  = \int_{\mathcal{P}(\mathcal{H})} q(Z) \mathcal{O}(Z) dV_{FS}
  ~,\label{eq:gqs}
\end{align}
where $\mathcal{O} \in \mathcal{A}$, $q(Z) \geq 0$ is the
normalized distribution associated with functional $P$:
\begin{align*}
P_q[\mathbb{I}] = \int_{\mathcal{P}(\mathcal{H})}
  q(Z) dV_{FS}  = 1
  ~,
\end{align*}
and $P_q[\mathcal{O}] \in \mathbb{R}$.

In this way, pure states $\ket{\psi_0}$ are functionals with a Dirac-delta
distribution $p_0(Z) = \widetilde{\delta}\left[ Z - Z_0\right]$:
\begin{align*}
P_{0}[\mathcal{O}] &= \intP \widetilde{\delta}(Z-Z_0)\mathcal{O}(Z) dV_{FS} \\
  & = \mathcal{O}(Z_0)  = \bra{\psi_0}\mathcal{O}\ket{\psi_0}
  ~.
\end{align*}
$\widetilde{\delta}(Z-Z_0)$ is shorthand for a coordinate-covariant Dirac-delta in
arbitrary coordinates. In homogeneous coordinates this reads:
\begin{align*}
\widetilde{\delta}(Z - Z_0) \coloneqq \frac{1}{\sqrt{\det g_{FS}}}
  \prod_{\alpha=0}^{D-1} \delta(X - X_0) \delta(Y - Y_0)
  ~,
\end{align*}
where $Z = X + iY$. In $(p_\alpha,\nu_\alpha)$ coordinates this becomes simply:
\begin{equation}
\widetilde{\delta}(Z - Z_0) = \prod_{\alpha=1}^{D-1} 2\delta(p_\alpha - p_\alpha^0) \delta(\nu_\alpha - \nu_\alpha^0)
  ~,
\end{equation}
where the coordinate-invariant nature of the functionals $P_q[\mathcal{O}]$ is
now apparent.

In this way, too, mixed states:
\begin{align*}
\rho = \sum_{j}\lambda_j \ket{\lambda_j}\bra{\lambda_j}
\end{align*}
are convex combinations of these Dirac-delta functionals:
\begin{align*}
q_{\mathrm{mix}}(Z) = \sum_{j}\lambda_j \widetilde{\delta}(Z-Z_j)
  ~. 
\end{align*}
Thus, expressed as functionals from observables to the real line, mixed states
are:
\begin{align}
P_{\mathrm{mix}}\left[ \mathcal{O}\right]
  & = \sum_{j} \lambda_j \bra{\lambda_j}\mathcal{O}\ket{\lambda_j}
  ~.
\label{eq:Functional}
\end{align}

Equipped with this formalism, one identifies the distribution $q(Z)$ as a
system's \emph{geometric quantum state}. This is the generalized notion of
quantum state we develop in the following.

A simple example of an ensemble that is neither a pure nor a mixed state is
the \emph{geometric canonical ensemble}:
\begin{align*}
q(Z) = \frac{1}{Q_\beta} e^{-\beta h(Z)}
  ~,
\end{align*}
where:
\begin{align*}
  Q_\beta & = \int dV_{FS} e^{-\beta h(Z)} ~, \\
  h(Z) & = \bra{\psi(Z)}H\ket{\psi(Z)} ~,
\end{align*}
and $H$ is the system's Hamiltonian operator. This state was previously
considered in Refs. \cite{Brody1998,Brody2016}. Reference \cite{Anza20b}
investigated its potential role in establishing a quantum foundation of
thermodynamics that is an alternative to that based on Gibbs ensembles and von
Neumann entropy. Moreover, it showed that the geometric ensemble genuinely 
differs from the Gibbs ensemble. This realization provides a concrete path 
to testing the experimental consequences of geometric quantum states.

\paragraph*{Density matrix.}
The connection between geometric quantum states and density matrices is
two-fold. On the one hand, when the distribution $q(Z)$ falls into one of the
two aforementioned cases---Dirac-deltas or finite convex combinations of
them---the present formalism is equivalent to the standard one. However, not
all functionals fall into the Dirac-delta form. Given this, $q(Z)$ is clearly a
more general notion of a quantum system's state.

On the other hand, given an arbitrary distribution $q(Z)$, there is a unique
density matrix $\rho^{q}$ associated to $q$:
\begin{align}
\rho^q_{\alpha \beta} & = P_q[Z^\alpha \overline{Z}^\beta] \nonumber \\
  & = \intP dV_{FS} \, q(Z)  \, Z^\alpha \overline{Z}^\beta
  ~.
\label{eq:densitymatrix}
\end{align}
Owing it to the fact that all POVMs are represented by real and quadratic
functions on $\mathcal{P}(\mathcal{H})$, recall Eq. (\ref{eq:GQM_POVMs}), they
are sensitive to $q(Z)$ via $\rho^q$. Therefore, if two
distributions $q_1$ and $q_2$ induce the same density matrix $\rho^{q_1} =
\rho^{q_2}$, then all POVMs produce the same outcomes.

A well-known consequence of this fact is that two density matrices with the
same barycenter are considered equal, even if they describe experiments with
different physical configurations. In these cases, the statistics of POVM
outcomes are described by the same density matrix. Note that this statement
does not mean that the two systems are in the same \emph{state}. Rather, it
means that there is no POVM on the system that distinguishes between $q_1$ and
$q_2$.

\begin{figure}[h]
\centering
\includegraphics[width=\columnwidth]{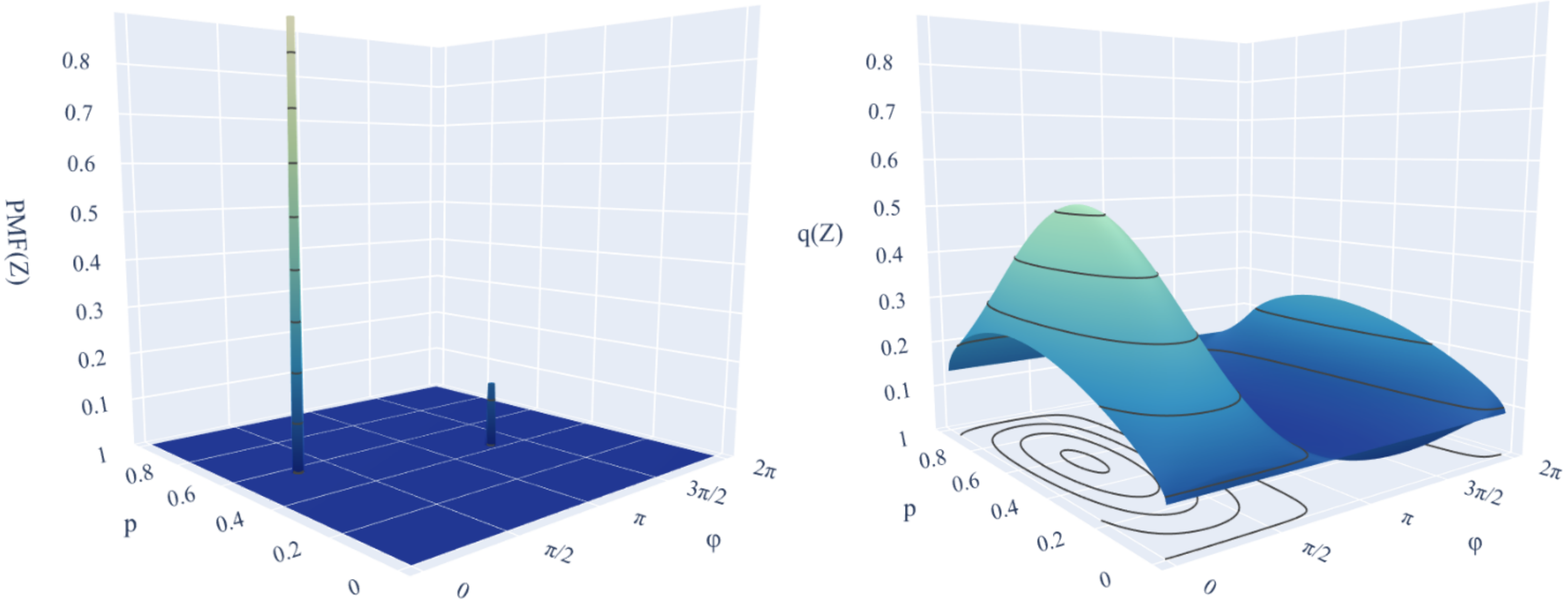}
\caption{Geometric quantum states in \emph{(probability,phase)} coordinates
	$(p,\phi)$ of $\mathbb{C}P^1$: (Left) Density matrix ``state'' $q_1$ is the
	convex sum of two Dirac delta-functions, centered on the eigenvectors
	$(p_+,\phi_+) = (0.568,0.983) $ and $(p_-, \phi_-) = (0.432,4.124)$ of
	density matrix $\rho$. (Right) Geometric quantum state $q_2$ differs
	markedly: A smooth distribution across the entire pure-state manifold
	$\mathbb{C}P^1$. However, $q_1$ and $q_2$ have the same density matrix
	$\rho_{q_1} = \rho_{q_2} = \rho$, where $\rho_{00} = 1- \rho_{11} = 0.45$,
	$\rho_{01} = \rho_{10}^{*} = 0.2 - 0.3 i$. $\rho_{\pm}$ are the eigenvalues
	of the density matrix: $\rho_+ = 0.864$ and $\rho_{-}=0.136$. Geometric
	quantum state $q_2$'s structure is only sparsely reflected in the
	density-matrix ``state'' $q_1$.
	}
\label{fig:Example} 
\end{figure}

To emphasize, consider the example of two geometric quantum states, $q_1$ and
$q_2$, with very different characteristics:
\begin{align*}
q_1(Z) & = \frac{1}{Q} e^{-\frac{1}{2} \overline{Z}\rho^{-1}Z } \\
q_2(Z) & = 0.864~\tilde{\delta}(Z - Z_{+}) + 0.136~\tilde{\delta}(Z - Z_{-})
  ~,
\end{align*}
where $Q = \int_{\mathbb{C}P^1} dV_{FS} e^{-\frac{1}{2}Z \rho^{-1}
\overline{Z}}$, $Z_+ = (0.657,0.418 + i 0.627)$, and $Z_- =
(0.754,-0.364-i0.546)$. However, states $q_1$ and $q_2$ have same density
matrix $\rho$ ($\rho_{00} = 0.45 = 1- \rho_{11}$ and $\rho_{01} = 0.2 - i 0.3
= \rho_{10}^{*}$) and so the same POVM outcomes. From Fig.\ref{fig:Example}
one appreciates the profound difference between $q_1$ and $q_2$, despite the
equality of their POVM statistics.

This is particularly important for quantum information processing where one
encounters long-range and long-lived correlational and mechanistic demands.
Quantum computing immediately comes to mind. There, one is not only interested
in measurement outcomes, but also in predicting and understanding how a quantum
system evolves under repeated external manipulations imposed by complex control
protocols.

\paragraph*{State manipulation.}
The following shows that the geometric formalism arises quite naturally when a
discrete quantum system interacts and develops entanglement with a continuous
one. Imagine a protocol controlling a system's continuous degrees of freedom to
manipulate discrete ones that store a computation's result. As a physical
reference, consider quantum particles with a given number of discrete degrees
of freedom (e.g., spin), confined to a region $\mathcal{R} \subseteq
\mathbb{R}^3$. The results we derive do not depend on this choice, since the
technical methods straightforwardly extend to other systems where continuous
and discrete degrees of freedom are mixed. A helpful illustration is
intra-particle entanglement \cite{Pasini2020}, that couples position and spin
degrees of freedom to create entangled states. In this way, one manipulates the spin
by only acting on the positional degrees of freedom, possibly via a potential.

Consider a hybrid quantum system comprised of $N$ continuous degrees of freedom
and $M$ qudits that are the discrete ones. The entire system's Hilbert space
is:
\begin{align*}
\mathcal{H} = \mathcal{H}^c_N \otimes \mathcal{H}^d_M
  ~,
\end{align*}
where $\mathcal{H}_N^c$ hosts the continuous degrees of freedom and has
infinite dimension, while $\mathcal{H}^d_M$ hosts the discrete ones and has
dimension $d^M$. A basis for $\mathcal{H}_N^c$ is provided by
$\left\{\ket{\vec{x}}\right\}$, where $\vec{x} \in \mathcal{R} \subseteq
\mathbb{R}^N$ and a basis for $\mathcal{H}_M^d$ is
$\left\{\ket{s}\right\}_{s=0}^{d^M-1}$. Thus, a generic state is:
\begin{align*}
\ket{\psi} = \int_{\mathcal{R}} d\vec{x} \sum_{s} \psi_s(\vec{x})\ket{\vec{x}} \ket{s}
  ~,
\end{align*}
where $\vec{x}$ is a dimensionless counterpart of the physical continuous
degrees of freedom, achieved by multiplying its value by appropriate physical
quantities. So, the measure $d\vec{x}$ has no physical dimension. For an
electron in a box, for example, this is achieved by renormalizing with the box's
total volume.

The following theorem establishes that this can be done constructively.

\begin{theorem}
\label{MainTheo}
Any state $\ket{\psi} \in \mathcal{H}$ can be written as:
\begin{align}
\ket{\psi} = \int_{\mathcal{R}}d\vec{x} f(\vec{x}) \ket{x}\ket{q(\vec{x})} \label{eq:theorem}
  ~,
\end{align}
where $f(\vec{x})$ is such that $\int_{\mathcal{R}}d\vec{x} |f(\vec{x})|^2 = 1$
and $\ket{q(\vec{x})}$ is a parametrized state of the discrete degrees of
freedom:
\begin{align*}
\ket{q(\vec{x})}
  = \sum_{s=0}^{d^M-1} \sqrt{p_s(\vec{x})} e^{i\phi_s(\vec{x})} \ket{s}
  ~,
\end{align*}
where $\left\{ p_s(\vec{x}),\phi_s(\vec{x})\right\}_s$ is a set of $2(d^M-1)$
real functions such that $\sum_{s=0}^{d^M-1} p_s(\vec{x}) = 1$,
$\phi_s(\vec{x}) \in [0,2\pi]$, and $\left\{\ket{s}\right\}_{s=0}^{d^M-1}$ is a
basis on $\mathcal{H}_M^d$.
\end{theorem}

(The Supplementary Material gives the proof.)
Equation (\ref{eq:theorem})'s state parametrization preserves key information
about the continuous degrees of freedom, namely $|f(\vec{x})|^2$, when working
with the discrete degrees of freedom. Indeed, the partial trace over the
continuous degrees of freedom yields:
\begin{align*}
\rho = \int_{\mathcal{R}} d\vec{x}
  \left\vert f(\vec{x})\right\vert^2 \ket{q(\vec{x})}\!\!\bra{q(\vec{x})}
  ~.
\end{align*}

Continuing, given an observable $\mathcal{O}$ with support only on
$\mathcal{H}^d_M$, we have:
\begin{align*}
\MV{\mathcal{O}}
  = \Tr \rho \mathcal{O} = \int_{\mathcal{R}} d\vec{x}
  \left\vert f(\vec{x})\right\vert^2 \mathcal{O}({q(\vec{x})})
  ~,
\end{align*}
where $\mathcal{O}({q(\vec{x})}) = \bra{q(\vec{x})}\mathcal{O}
\ket{q(\vec{x})}$. Comparing with Eq. (\ref{eq:gqs})
one realizes that the functions $\left\{p_s(\vec{x}),\phi_s(\vec{x})\right\}$ provide an $\vec{x}$-dependent embedding of 
$\mathcal{R} \subseteq \mathbb{R}^N$ onto $\mathbb{C}P^n$, with $n=d^M-1$, or
a submanifold, via:
\begin{align*}
\Phi : \vec{x} \to \Phi(\vec{x}) = Z(\vec{x})
  ~,
\end{align*}
where:
\begin{align*}
Z = (Z^0,\ldots,Z^n)
  ~,
\end{align*}
with $Z^\alpha(\vec{x}) = \sqrt{p_\alpha(\vec{x})}e^{i\phi_\alpha(\vec{x})}$.
Thus, letting $\mathcal{R}^{*} = \Phi(\mathcal{R})$, we obtain:
\begin{align*}
\int_{\mathcal{R}} d\vec{x} \left\vert f(\vec{x})\right\vert^2 \mathcal{O}({q(\vec{x})}) = \int_{\mathcal{R}^{*}} dV_{FS} \,\, q(Z) \mathcal{O}(Z)
  ~,
\end{align*}
where:
\begin{align*}
q(Z) = \frac{\left\vert \det D\Phi (Z)
  \right\vert }{\sqrt{\det g_{FS}}} \left\vert f(\Phi^{-1}(Z)) \right\vert^2
  ~.
\end{align*}
Here, $D\Phi$ denotes the Jacobian of the transformation $\Phi$ and $g_{FS}$ is
the Fubini-Study metric tensor and we assume the transformation is invertible.
Generalizing to cases in which $\Phi^{-1}$ is not invertible, due to the fact
that different $\vec{x}$ might yield the same $\left(
p_s(\vec{x}),\phi_s(\vec{x})\right)$, is left to future efforts.

Let's illustrate with a familiar system: an electron in a 2D rectangular box
$\mathcal{R} = [x_0,x_1] \times [y_0,y_1]$. In this case, $M=1$ and $d=2$ so
that we have $f(x,y)$, $\left\{p_s(x,y), \phi_s(x,y)\right\}_{s=0,1}$. This
amounts to:
\begin{align*}
\int_{x_0}^{x_1} \!\!\! dx & \int_{y_0}^{y_1} \!\!\!dy
  |f(x,y)|^2 \mathcal{O}(q(x,y)) \\
  & = \frac{1}{2}\int_0^1 \!\!\!dp \int_0^{2\pi}\!\!\!\!\!d\phi \, q(Z(p,\phi)) \, O(Z(p,\phi))
  ~,
\end{align*}
where, for example, $p_0(x,y) = 1 - p_1(x,y)$, $p_1(x,y) = \frac{x -
x_0}{x_1-x_0}$, $\phi_0(x,y) = 0$, and $\phi_1(x,y) = 2\pi \frac{y -
y_0}{y_1-y_0}$.

In short, a generic quantum state $\ket{\psi}$ of the whole system uniquely
defines a distribution $q(Z)$ on the manifold of pure states
$\mathcal{P}(\mathcal{H}_{M}^d) = \mathbb{C}P^{d^M-1}$. The correspondence is
not one-to-one as knowledge of $q(Z)$ does not allow recovering the entire
state. The missing part is $\theta_0(\vec{x})$, the phase of $f(\vec{x})$.
However, it does circumscribe the possible states as it fixes the shape of the
probability distribution of the continuous variables $|f(\vec{x})|^2$.

Note how the embedding functions $p_s(\vec{x})$ and $\phi_s(\vec{x})$ play a
key role in determining whether we can cover the whole $\mathbb{C}P^n$ or just
a submanifold. Consider the conditions that guarantee the two extreme cases are
covered: full covering of $\mathbb{C}P^n$ and covering of tensor product states
only $\mathbb{C}P^{d-1} \otimes \ldots \otimes \mathbb{C}P^{d-1}$. In the first
case, $\mathbb{C}P^n$ is a complex manifold that requires $2n$ independent real 
coordinates to be completely covered. For $M$ qudits this means:
\begin{align*}
M & \leq M^{\mathrm{Full}}_{\max} \\
  & = \frac{\log \left(N/2 + 1\right)}{\log d}
  ~.
\end{align*}
Instead, if we need to cover only the submanifold of tensor product states, the
number of qudits we can control with $N$ continuous degrees of freedom is much
larger:
\begin{align*}
M & \leq M^{\mathrm{Prod}}_{\max} \\
  & = \frac{N}{2(d-1)}
  ~.
\end{align*}
Most cases fall in between. And so, the number of qudits controllable with $N$
continuous variables is $M \in \left[M^{\mathrm{Full}}_{\max},
M^{\mathrm{Prod}}_{\max}\right]$.

\paragraph*{Thermodynamic framework.}
Another setting in which the geometric formalism arises naturally is quantum
thermodynamics. There, one is often interested in modeling the behavior of a
small system in a thermal environment. For modest-sized environments one can
naively treat the system and environment as isolated and then simulate its
evolution. As the environment's size grows, this quickly becomes infeasible.
Nonetheless, as we now show, the geometric formalism allows appropriately
writing the system's reduced density matrix in a way that retains much of the
information about the environment. This can be done due to Thm. \ref{MainTheo}.

Consider a large quantum system consisting of $M$ qudits split in two
asymmetric parts. Call the small part with $N_S$ qudits the ``system'' and
let the rest be the ``environment'' with $N_E = M-N_S$ qudits. A generic state
of the entire system $\mathcal{H}_S \otimes \mathcal{H}_E$ is $\ket{\psi_{SE}}
= \sum_{k=0}^{d_S-1}\sum_{\alpha=0}^{d_E-1}
\psi_{k\alpha}\ket{s_k}\ket{e_\alpha}$, where $\left\{\ket{s_k}\right\}_k$ and
$\left\{\ket{e_\alpha}\right\}_k$ are bases for $\mathcal{H}_S$ and
$\mathcal{H}_E$, respectively.

Given $\ket{\psi_{SE}}$, it is not too hard to see that the system's (reduced) state is:
\begin{align}
\rho^S = \sum_{\alpha=1}^{d_E} p_\alpha^S \ket{\chi_\alpha^S} \bra{\chi_\alpha^S}\label{eq:reducedS}
  ~,
\end{align}
where:
\begin{align*}
p_\alpha^S = \sum_{k=0}^{d_S-1} \left\vert \psi_{k\alpha} \right\vert^2
  ~,
\end{align*}
and
\begin{align*}
\ket{\chi_\alpha^S} = \frac{1}{\sqrt{p_\alpha^S}} \sum_{k=0}^{d_S-1} \psi_{k\alpha}\ket{s_k}
  ~.
\end{align*}

In numerical analysis one often retains only the $d_S \times d_S$ matrix
elements of $\rho^S$ in a certain basis. However, this erases the functional
information about the environment. Instead, the latter can be recovered from
$\left\{ p_\alpha^S, \ket{\chi_\alpha^S}\right\}$ as:
\begin{align*}
\left(\rho^E\right)_{\alpha\beta}
  = \sqrt{p_\alpha^S p_\beta^S}\braket{\chi_\alpha^S}{\chi_\beta^S}
  ~.
\end{align*}
As $d_E$ grows, retaining this information as a set of probabilities and states
quickly becomes unrealistic.

However, the same information can be effectively encoded by switching to a
geometric description. Indeed, at finite $d_E$, $\rho^S$ becomes:
\begin{align*}
p^S_{d_E}(Z) = \sum_{\alpha=1}^{d_E}p_\alpha^E
  \tilde{\delta}\left[ Z - Z(\chi_\alpha^S)\right]
\end{align*}
and the thermodynamic limit is conveniently handled with:
\begin{align*}
p_\infty(Z) = \lim_{d_E \to \infty} \sum_{\alpha=1}^{d_E}
  p_\alpha^E \tilde{\delta}\left[ Z - Z(\chi_\alpha^S)\right]
  ~.
\end{align*}
Here the limit is performed, as usual, by keeping finite the average energy density $\lim_{d_E \to \infty} \MV{H}/(N_S + N_E) = \epsilon$.

In this way, the geometric formalism emerges naturally in a quantum
thermodynamics. In the limit of large environments, one simply cannot keep
track of exactly how an environment generates the ensemble of our system under
study and so switch to a probabilistic description. Helpfully, the geometric
formalism efficiently controls this. See also Ref. \cite{Anza20b} for an
expanded exploration of the geometric formalism in quantum thermodynamics.

Before proceeding, though, let's highlight an interesting discrepancy between
the two applications presented. In the thermodynamic setting, knowledge of the
ensemble $\left\{ p_\alpha^S, \ket{\chi_\alpha^S}\right\}$ allows fully
recovering the global pure state $\ket{\psi_{SE}}$. Indeed, it is easy to see
that:
\begin{align*}
\psi_{k\alpha} = \sqrt{p_\alpha^S} \braket{s_k}{\chi_\alpha^S}
  ~.
\end{align*}
Substituting this into the pure state $\ket{\psi_{SE}}$, we obtain a Schmidt-like
decomposition in which the common label runs over the dimension of the 
environment's Hilbert space:
\begin{align*}
\ket{\psi_{SE}} = \sum_{\alpha=0}^{d_E-1}
  \sqrt{p_\alpha^S} \ket{\chi_\alpha^S}\ket{e_\alpha}
  ~.
\end{align*}
The price paid for the decomposition is that the states $\ket{\chi_\alpha^S}$
are not orthogonal. However, we gain a more detailed description of our
system's state. As we can see, here the challenge of recovering
$\ket{\psi_{SE}}$ disappears thanks to $p_\alpha^S \in \mathbb{R}$. We comment
on this discrepancy with the other case shortly.

\paragraph*{Discussion.}
Standard quantum mechanics' concept of state is the density matrix. However,
while density matrices provide a complete account of POVM statistics, they are
not in one-to-one correspondence with the ensembles that generated them. This
is a well-known fact that underlies the freedom in writing a decomposition of
the density matrix in terms of probabilities and pure states.  All such
decompositions yield the same POVM statistics, but they are not physically
equivalent since they are realized in physically different ways.

From a purification perspective \cite{Wilde2017}, the physical information
about an ensemble's realization can always be thought of as coming from a
larger system that is in a pure state. While the additional information about
how the ensemble is realized is not relevant for the measurement statistics on
our system, it does provide a much richer description. It preserves part (if not all) 
of the structural information about how the system's POVM statistics result from
interactions with its surroundings.
 
Geometric quantum mechanics and its concept of geometric quantum state provide
a framework that allows retaining such information. This yields a richer
picture of the system's state which goes beyond the system's POVM statistics, 
taking into account the physical way in which an ensemble has been realized.
The geometric formalism's benefits emerge in at least two important cases:
(i) Hybrid continuous-discrete systems, e.g., electrons or other particles with
spin or other discrete degrees of freedom, and (ii) the thermodynamic setting
of a system in contact with a large environment.

The geometric formalism directly handles the continuous nature of hybrid
systems and the large number of degrees of freedom in thermodynamics. And, it
does so in a fairly simple way. This allows working with the full geometric
quantum state, thus retaining the structural information about how the ensemble
is generated. While the two applications considered are similar, a crucial
difference does appear. If we assume a finite environment, knowledge of the
geometric quantum state of our system is sufficient to recover the global pure
state of system and environment. This does not occur for a hybrid
discrete-continuous system, where knowledge of the geometric quantum state does
not allow inferring the phase $\theta_0(\vec{x})$ of $f(\vec{x})$. Notably,
fully recovering the overall pure state, whose physical relevance can be argued
on the ground of continuity with the finite-dimensional case, effectively
translates into a $U(1)$ gauge principle on the overall system. The requirement
that states differing from a local phase are physically
equivalent---$\psi_s(\vec{x}) \sim
e^{i\varphi(\vec{x})}\psi_s(\vec{x})$---turns into a sufficient condition for
recovering the global state from the geometric quantum state since, in this
case, one can always choose $f(\vec{x}) \in \mathbb{R}$. We leave exploring the
connection between recovering the global pure state from a local geometric
quantum state and a gauge principle for a future investigation.

\paragraph*{Conclusion.}
Geometric quantum mechanics is an alternative to the standard vector-based
formalism. We introduced and then explored the concept of \emph{geometric
quantum state} $p(Z)$ as a probability distribution on the manifold of pure
states, inspired by the statistics of chaotic attractors from the theory of
dynamical systems or, more appropriately, its Sinai-Bowen-Ruelle measures
\cite{Eckmann1985}. This characterization of a quantum state accounts for the
fact that singling out the density matrix as the sole descriptor of a quantum
system's state entails ignoring how an ensemble is physically realized. While
this does not have consequences for POVM statistics, in concrete situations the
information about the ensemble realization can be key to accurate modeling.
Reference \cite{Anza20c} gives an example. That said, density matrices can be
readily computed as quadratic averages from $p(Z)$ via Eq.
(\ref{eq:densitymatrix}).

We explored the physical relevance of geometric quantum states via an open
quantum system in which a (finite) system under study is in contact with a
larger environment and the joint state is assumed to be pure. In this
thermodynamic setting, portions of the structural information about the joint
pure state is directly preserved in the geometric quantum state of the smaller
system under study. The result is a markedly richer picture of the system's
state---a picture that goes substantially beyond the density matrix and its
POVM statistics.

\section*{Acknowledgments}
\label{sec:acknowledgments}

F.A. thanks Marina Radulaski, Davide Pastorello, and Davide Girolami for
discussions on the geometric formalism of quantum mechanics. F.A. and J.P.C.
thank Dhurva Karkada for his help with the example, David Gier, Samuel Loomis, 
and Ariadna Venegas-Li for helpful discussions and the Telluride Science Research 
Center for its hospitality during visits.  This material is based upon work supported 
by, or in part by, a Templeton World Charity Foundation Power of Information
Fellowship, FQXi Grant FQXi-RFP-IPW-1902, and U.S. Army Research Laboratory 
and the U. S. Army Research Office under contracts W911NF-13-1-0390 and
W911NF-18-1-0028.

\makeatletter
\newcommand{\manuallabel}[2]{\def\@currentlabel{#2}\label{#1}}
\makeatother

\clearpage
\appendix
\onecolumngrid

\pagestyle{empty}

\begin{center}
\large{Supplementary Materials}\\
\vspace{0.1in}
\emph{\ourTitle}\\
\vspace{0.1in}
{\small
Fabio Anza and James P. Crutchfield
}
\end{center}

\section{The Search for Quantum States}
\label{sm:QStates}

In those domains of the physical sciences that concern the organization and
evolution of systems, a common first task is to determine a system's distinct
configurations or \emph{effective states}. Ultimately, this turns on what
questions there are to answer. One goal is prediction---of properties or
behaviors. And, in this, quantum mechanics stands out as a particularly telling
arena in which to define effective states.

The very early history of its development can be construed partially as
attempts to answer this question, from de Broglie's \emph{phase-waves}
\cite{Brog25a} and Schrodinger's \emph{wave functions} \cite{Schr26a} to von
Neumann's \emph{statistical operators} in Refs. \cite{Neum27a} and \cite[Chap.
IV]{Neum32a}, later labeled \emph{density matrices} by Dirac
\cite{Dira29a,Dira30b,Dira31a}. And, these were paralleled by Heisenberg's
``operational'' \emph{matrix mechanics} that focused on experimentally
accessible observables and so avoided imputing internal, hidden structure
\cite{Heis25a}.

The abiding challenge is that effective states are almost always inferred
indirectly and through much trial and error. Quantum mechanics heightens the
challenge greatly due to its foundational axiom that the detailed, microscopic,
and fundamental degrees of freedom cannot be directly and completely measured
\emph{in principle}. The main text revisits this perennial question, What is a
quantum state?

\section{Theorem \ref{MainTheo}: Proof}
\label{sm:MainTheo}

In this Appendix we give the detailed proof of Theorem \ref{MainTheo} in the main text. 
Let's first restate the setup of the theorem.

Consider a hybrid quantum system comprised of $N$ continuous degrees of freedom
and $M$ qudits that are the discrete ones. The entire system's Hilbert space
is:
\begin{align*}
\mathcal{H} = \mathcal{H}^c_N \otimes \mathcal{H}^d_M
  ~,
\end{align*}
where $\mathcal{H}_N^c$ hosts the continuous degrees of freedom and has
infinite dimension, while $\mathcal{H}^d_M$ hosts the discrete ones and has
dimension $d^M$. A basis for $\mathcal{H}_N^c$ is provided by
$\left\{\ket{\vec{x}}\right\}$, where $\vec{x} \in \mathcal{R} \subseteq
\mathbb{R}^N$ and a basis for $\mathcal{H}_M^d$ is
$\left\{\ket{s}\right\}_{s=0}^{d^M-1}$. Thus, a generic state is:
\begin{align*}
\ket{\psi} = \int_{\mathcal{R}} d\vec{x} \sum_{s} \psi_s(\vec{x})\ket{\vec{x}} \ket{s}
  ~,
\end{align*}
where $\vec{x}$ is a dimensionless counterpart of the physical continuous
degrees of freedom, achieved by multiplying its value by appropriate physical
quantities. So, the measure $d\vec{x}$ has no physical dimension.

\emph{{\bf Theorem \ref{MainTheo}.}
Any state $\ket{\psi} \in \mathcal{H}$ can be written as:
\begin{align*}
\ket{\psi} = \int_{\mathcal{R}}d\vec{x} f(\vec{x}) \ket{x}\ket{q(\vec{x})}
  ~,
\end{align*}
where $f(\vec{x})$ is such that $\int_{\mathcal{R}}d\vec{x} |f(\vec{x})|^2 = 1$
and $\ket{q(\vec{x})}$ is a parametrized state of the discrete degrees of
freedom:
\begin{align*}
\ket{q(\vec{x})}
  = \sum_{s=0}^{d^M-1} \sqrt{p_s(\vec{x})} e^{i\phi_s(\vec{x})} \ket{s}
  ~,
\end{align*}
where $\left\{ p_s(\vec{x}),\phi_s(\vec{x})\right\}_s$ is a set of $2(d^M-1)$
real functions such that $\sum_{s=0}^{d^M-1} p_s(\vec{x}) = 1$,
$\phi_s(\vec{x}) \in [0,2\pi]$, and $\left\{\ket{s}\right\}_{s=0}^{d^M-1}$ is a
basis for $\mathcal{H}_M^d$.
}

\emph{Proof}: The proof is constructive. Given an arbitrary
$\left\{ \psi_s(\vec{x}) \right\}_s$, we can always find the set of functions
$f(\vec{x})$, $p_s(\vec{x})$, and $\phi_s(\vec{x})$. The converse holds
trivially: Given these functions one can always compute the $\left\{
\psi_s(\vec{x}) \right\}_s$. The set of transformations that maps one
parametrization into the other is:
\begin{align*}
\phi_s(\vec{x}) = \theta_s(\vec{x}) - \theta_0(\vec{x})
  ~,
\end{align*}
where:
\begin{align*}
e^{i\theta_s(\vec{x})} = \frac{\psi_s(\vec{x})}{|\psi_s(\vec{x})|}
  ~.
\end{align*}
Moreover:
\begin{align*}
f(\vec{x}) & = \sqrt{\sum_{s=0}^{d^M-1} \vert \psi_s(\vec{x}) \vert^2}
  \, e^{i \theta_0(\vec{x})} ~\text{and} \\
p_s(\vec{x}) & \coloneqq
  \frac{\vert \psi_s(\vec{x})\vert^2}{\sum_{l=0}^{d^M-1}\left\vert \psi_l(\vec{x})\right\vert^2}
  ~,
\end{align*}
It is easy to see how normalization of $|f(\vec{x})|^2$ and of $p_s(\vec{x})$ emerges from the definitions:
\begin{align*}
\int_{\mathcal{R}} d\vec{x} \left\vert f(\vec{x})\right\vert^2
  & = \int_{\mathcal{R}} d\vec{x}
  \sum_{s=0}^{d^M-1} \left\vert \psi_s(\vec{x}) \right\vert^2 \\
  & = 1 ~,\\
\sum_{s=0}^{d^M-1} p_s(\vec{x})
  & = \sum_{s=0}^{d^M-1}
  \frac{\vert \psi_s(\vec{x})\vert^2}{\sum_{l=0}^{d^M-1}\left\vert
  \psi_l(\vec{x})\right\vert^2} \\
  & = 1 ~,~\text{and} \\
\vert e^{i\phi_s(\vec{x})}\vert^2
  & = \frac{|\psi_s(\vec{s})|}{|\psi_s(\vec{x})|} \\
  & = 1
  ~.
\end{align*}
The latter gives $\phi_s(\vec{x}) \in [0,2\pi]$.

With these definitions we obtain:
\begin{align*}
e^{i\phi_s(\vec{x})} f(\vec{x}) \sqrt{p_s(\vec{x})}
  & = \sqrt{|\psi_s(\vec{x})|^2} e^{i\theta_s(\vec{x})} \\
  & = \psi_s(\vec{x})
  ~.
\end{align*}
This in turn gives the desired result:
\begin{align*}
\ket{\psi} & = \int_{\mathcal{R}} d\vec{x} \sum_{s} \psi_s(\vec{x})\ket{\vec{x}} \ket{s} \\
& = \int_{\mathcal{R}} d\vec{x} f(\vec{x}) \ket{\vec{x}}\sum_{s} e^{i\phi_s(\vec{x})}   \sqrt{p_s(\vec{x})}  \ket{s} \\
& = \int_{\mathcal{R}}d\vec{x} f(\vec{x}) \ket{x}\ket{q(\vec{x})}
  ~.
\end{align*}

\bibliographySM{library}

\end{document}